\documentclass[preprint,prb,aps,floats,amssymb,showpacs,draft,%
floatfix,superscriptaddress]{revtex4} 
\usepackage{psfig}

\begin{document}

\title{ 
Magnetic-glassy multicritical behavior of the three-dimensional $\pm J$ Ising model
} 

\author{Martin Hasenbusch} 
\affiliation{ 
Dipartimento di Fisica dell'Universit\`a di Pisa and INFN, Pisa,
  Italy.  } 
\author{Francesco Parisen Toldin} 
\affiliation{ 
Scuola Normale Superiore and INFN, Pisa, Italy.  } 
\author{Andrea Pelissetto} 
\affiliation{Dipartimento di Fisica
  dell'Universit\`a di Roma ``La Sapienza" and INFN, Roma, Italy.}
\author{Ettore Vicari} 
\affiliation{ 
Dipartimento di Fisica dell'Universit\`a di Pisa and INFN, Pisa, Italy.  } 

\date{\today}

\begin{abstract}
  We consider the three-dimensional $\pm J$ model defined on a simple cubic
  lattice and study its behavior close to the multicritical Nishimori point
  where the paramagnetic-ferromagnetic, the paramagnetic-glassy, and the
  ferromagnetic-glassy transition lines meet in the $T$-$p$ phase diagram ($p$
  characterizes the disorder distribution and gives the fraction of
  ferromagnetic bonds).  For this purpose we perform Monte Carlo simulations
  on cubic lattices of size $L\le 32$ and a finite-size scaling analysis of
  the numerical results.  The magnetic-glassy multicritical point is found at
  $p^*=0.76820(4)$, along the Nishimori line given by $2p-1={\rm Tanh}(J/T)$.
  We determine the renormalization-group dimensions of the operators that
  control the renormalization-group flow close to the multicritical point,
  $y_1 = 1.02(5)$, $y_2 = 0.61(2)$, and the susceptibility exponent $\eta =
  -0.114(3)$.  The temperature and crossover exponents are $\nu=1/y_2=1.64(5)$
  and $\phi=y_1/y_2 = 1.67(10)$, respectively.  We also investigate the
  model-A dynamics, obtaining the dynamic critical exponent $z = 5.0(5)$.
\end{abstract}



\maketitle


\section{Introduction}

The $\pm J$ Ising model provides an interesting
theoretical laboratory to study the effects of quenched random disorder and
frustration in Ising systems. It is defined by the lattice Hamiltonian
\begin{equation}
{\cal H} = - \sum_{\langle xy \rangle} J_{xy} \sigma_x \sigma_y,
\label{lH}
\end{equation}
where $\sigma_x=\pm 1$, the sum is over the nearest-neighbor sites of a simple
cubic lattice, and the exchange interactions $J_{xy}$ are uncorrelated
quenched random variables, taking values $\pm J$ with probability distribution
\begin{equation}
P(J_{xy}) = p \delta(J_{xy} - J) + (1-p) \delta(J_{xy} + J). 
\label{probdis}
\end{equation}
In the following we set $J=1$ without loss of generality. 
For $p=1$ we recover the standard ferromagnetic
Ising model, while for $p=1/2$ we obtain the bimodal Ising
spin-glass model.  
The $\pm J$ Ising model is a simplified model \cite{EA-75} for
disordered spin systems showing glassy behavior in some region of their
phase diagram, such as Fe$_{1-x}$Mn$_x$TiO$_3$ and
Eu$_{1-x}$Ba$_x$MnO$_3$, see,
e.g., Refs.~\onlinecite{IATKST-86,GSNLAI-91,NN-07}.  The random nature
of the short-ranged interactions is mimicked by nearest-neighbor random
bonds.

\begin{figure*}[tb]
\centerline{\psfig{width=9truecm,angle=0,file=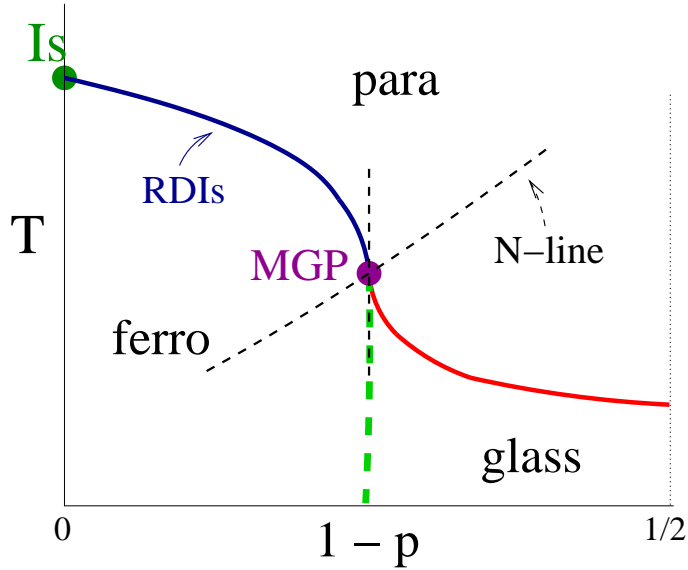}}
\vspace{2mm}
\caption{
Phase diagram of the three-dimensional $\pm J$ Ising model in the $T$-$p$
plane. The phase diagram is symmetric for $p\rightarrow 1-p$.}
\label{phdia}
\end{figure*}

The $T$-$p$ phase diagram of the three-dimensional $\pm J$ Ising model is
sketched in Fig.~\ref{phdia} for $1\ge p \ge 1/2$ (it is symmetric for
$p\rightarrow 1-p$).  The high-temperature phase is paramagnetic for any $p$.
The nature of the low-temperature phase depends on the value of $p$: it is
ferromagnetic for small values of $1-p$, while it is glassy with vanishing
magnetization for sufficiently large values of $1-p$.  The paramagnetic and
low-temperature ferromagnetic and glassy phases are separated by different
transition lines, which meet at a magnetic-glassy multicritical point (MGP)
located at $p^*, T^*$ and usually called Nishimori point.

The paramagnetic-ferromagnetic (PF) transition line starts from the
Ising transition at $p=1$ and extends up to the MGP at $p=p^*$. For
$p=1$ the transition belongs to the Ising universality class, while
for any $1>p>p^*$ it belongs to the randomly-dilute Ising (RDIs)
universality class,\cite{Hukushima-00,HPPV-07-pmj} characterized by
the magnetic critical exponents
\cite{HPPV-07,PV-02} $\nu_{f}=0.683(2)$ and $\eta_{f}=0.036(1)$. 
The Ising transition at $p=1$ is a multicritical point and, close to
it, for $0<1-p\ll 1$, one observes multicritical
behavior\cite{HPPV-07-pmj,CPPV-04,Aharony-76} with crossover exponent
$\phi=\alpha_{\rm Is}$, where\cite{CPRV-02} $\alpha_{\rm
Is}=0.1096(5)$ is the Ising specific-heat exponent. The
paramagnetic-glassy (PG) transition line starts from the MGP and
extends up to $p=1/2$.  A reasonable hypothesis is that the critical
behavior is independent of $p$ along the PG line, i.e. that a nonzero
average value $[J_{xy}]$ of the bond variables is irrelevant at the
glass transition, as found in mean-field models.\cite{meanfield}
Assuming this scenario, for any $1-p^* < p < p^*$ the PG transition
belongs to the same universality class as that of the bimodal Ising
spin glass model at $p=1/2$. Its critical behavior has been widely
investigated (see, e.g., Refs.~\onlinecite{KKY-06,KR-03} and
references therein) and it is characterized by the overlap exponents
$\nu_g\approx 2.4$ and $\eta_g\approx -0.4$.

As argued in Refs.~\onlinecite{GHDB-85,LH-88,LH-89}, the MGP is
located along the so-called Nishimori line~\cite{Nishimori-81,Nishimori-book} 
(N-line) defined by the relation
\begin{equation}
v\equiv {\rm Tanh} \beta = 2 p -1,
\label{nishline}
\end{equation}
where $\beta\equiv 1/T$, which allows us to define a Nishimori temperature
\begin{equation}
\beta_N(p) = {1\over T_N(p)} = {1\over2} \ln {p\over 1-p}
\end{equation}
for each value of $p$.
The $\pm J$ Ising model along the N-line presents several
interesting properties.  The internal energy has been computed
exactly along the N-line:\cite{Nishimori-81}
\begin{equation}
E_N(p) = {1\over V} [ \langle {\cal H} \rangle_{T_N(p)} ] = 6p-3,
\label{energy}
\end{equation}
where the angular parentheses and the brackets refer respectively
to the thermal average and to the quenched average over the bond
couplings $\{J_{xy}\}$.
Along the N-line several other remarkable relations
hold, such as~\cite{Nishimori-81} 
\begin{equation}
[\langle A_X \rangle ] = [\langle A_X \rangle^2],
\label{nishrel}
\end{equation}
where $A_X$ is an arbitrary product of spin variables $\sigma_x$,
and also~\cite{LH-88} 
\begin{equation}
G_{2i+1}(x) = G_{2i+2}(x) \qquad i=1,2,....
\label{Grel}
\end{equation}
where  $G_k(x) \equiv [ \langle \sigma_0 \,\sigma_x \rangle^k ]$.
As a consequence of Eq.~(\ref{Grel}),
the magnetic correlation function $G_1(x)$ and the overlap
correlation function $G_2(x)$ are equal along the N-line.
The N-line separates the regions where magnetic and
glassy fluctuations dominate.  Arguments based on local gauge
invariance \cite{GHDB-85,LH-88,LH-89} show that the MGP must be
located along the N-line, so that $T^* = T_N(p^*)$.
At the MGP, magnetic and
glassy fluctuations become critical simultaneously.  

At fixed $p$ an important inequality holds:\cite{Nishimori-81,KR-03}
\begin{equation}
| [ \langle \sigma_i \sigma_j \rangle_T ] | \le
[ | \langle \sigma_i \sigma_j \rangle_{T_N(p)} | ] ,
\label{ineq}
\end{equation}
where the subscripts indicate the temperature of the thermal average.
This relation shows that 
ferromagnetism can exist only in the region
$p>p^*$ and that the system is maximally magnetized along the 
N-line. Ref.~\onlinecite{Nishimori-86} (see also 
Refs.~\onlinecite{Kitatani-92,Nishimori-book}) also reports 
an argument that indicates that the ferromagnetic-glassy (FG)
transition line coincides with the line $p=p^*$, from $T=T^*$ to $T=0$.
This conjecture is contradicted by recent results for the 
two-dimensional $\pm J$ model \cite{WHP-03,AH-04,PHP-06}
and for three-dimensional random-plaquette gauge model,
\cite{WHP-03} which is the dual of the $\pm J$ model.
Violations are in any case quite small.
We mention that a mixed low-temperature phase,\cite{Kitatani-94}
in which ferromagnetism
and glass order coexist, is found in mean-field models~\cite{meanfield}
such as the infinite-range Sherrington-Kirkpatrick model.\cite{SK-75}
Its presence has been confirmed in the $\pm J$ Ising model defined on
Bethe lattices.\cite{CKR-05}  However, there is no evidence of this mixed
phase in the $\pm J$ Ising model on a cubic
lattice~\cite{Hartmann-99} and in related models.\cite{KM-02}
Nevertheless, the existence of such a mixed phase is still 
an open problem, as discussed in Ref.~\onlinecite{CKR-05}.

In this paper we consider the $\pm J$ model and perform Monte Carlo
(MC) simulations along the N-line close to the MGP.  By performing a
finite-size scaling (FSS) analysis, we locate the multicritical point
along the N-line, finding $p^*=0.76820(4)$.  We determine the
renormalization-group (RG) dimensions $y_1$ and $y_2$ of the relevant
operators that control the RG flow close to the MGP and the exponent
$\eta$ that gives the critical behavior of the magnetic and of the
overlap susceptibility.  We obtain $y_1 = 1.02(5)$, $y_2 = 0.61(2)$,
and $\eta = -0.114(3)$.  The temperature and crossover exponents are
$\nu=1/y_2=1.64(5)$ and $\phi=y_2/y_1 = 1.67(10)$ respectively.  We
also use our numerical results to estimate the dynamic critical
exponent $z$ that characterizes the model-A dynamics~\cite{HH-77} at
the MGP, i.e. a relaxational dynamics without conserved order
parameters.  We obtain $z = 5.0(5)$. Our results significantly improve
those obtained in previous
works.\cite{ON-87,Fisch-91,Singh-91,SA-96,MB-98,OI-98}

The paper is organized as follows.  In Sec.~\ref{sec2} we summarize
the theoretical results we need in our numerical analysis. In
Sec.~\ref{sec3} we report our numerical results.  We estimate the
position of the MGP and the critical exponents $y_1$, $y_2$, and
$\eta$ in Sec.~\ref{sec3.1}, while in Sec.~\ref{sec3.2} we give an
estimate of the exponent $z$ for the Metropolis dynamics we use, which
is a specific example of a relaxational dynamics without order
parameters (the so-called model-A dynamics). In Sec.~\ref{sec4} we
summarize our results.  In the Appendix we report some notations.

\section{Summary of theoretical results} \label{sec2}

In the absence of external fields, the critical behavior at the MGP is
characterized by two relevant RG operators. The
singular part of the free energy averaged over disorder 
in a volume of size $L$ can be written as
\begin{equation}
F_{\rm sing}(T,p,L) = L^{-d} f(u_1 L^{y_1}, u_2 L^{y_2}, \{u_i L^{y_i}\}),\quad i\ge 3,
\label{freeen}
\end{equation} 
where $y_1>y_2>0$, $y_i<0$ for $i\ge 3$, $u_i$ are the corresponding 
scaling fields, and $u_1 = u_2 = 0$ at the MGP.  In the infinite-volume
limit and neglecting subleading corrections, we have
\begin{equation}
F_{\rm sing}(T,p) = |u_2|^{d/y_2} f_\pm (u_1 |u_2|^{-\phi}), \qquad \phi=y_1/y_2>1,
\label{freeen2}
\end{equation} 
where the functions $f_\pm(x)$ apply to the parameter regions in which 
$\pm u_2 > 0$. Close to the MGP, all transition lines correspond to 
constant values of the product $u_1 |u_2|^{-\phi}$ and thus,
since $\phi > 1$, they are tangent to the line $u_1 = 0$.

The scaling fields $u_i$ are analytic functions of the model
parameters $T$ and $p$.  Using symmetry arguments, 
Refs.~\onlinecite{LH-88,LH-89} showed that one scaling axis is along
the N-line, i.e. that the N-line is either tangent to the 
line $u_1 = 0$ or to $u_2 = 0$. Since the N-line cannot be tangent 
to the transition lines at the MGP and these lines are tangent to 
$u_1 = 0$, the first possibility is
excluded. Thus, close to the MGP the N-line corresponds to $u_2 = 0$.
Thus, we identify\cite{LH-88,LH-89}
\begin{equation}
u_2=v-2p+1.
\label{u2sf}
\end{equation} 
As for the scaling axis $u_1 = 0$, 
$\epsilon\equiv 6-d$ expansion calculations predict it 
\cite{LH-89} to be parallel to the $T$ axis.
The extension of this result to $d=3$ suggests 
\begin{equation}
u_1=p-p^*.
\label{u1sf}
\end{equation} 
Note that, if Eq.~(\ref{u1sf}) holds, only the scaling field $u_2$ 
depends on the temperature $T$. We may then identify 
$\nu=1/y_2$ as the critical exponent associated with the temperature, and rewrite
Eq.~(\ref{freeen2}) as
\begin{equation}
F_{\rm sing}(T,p) = |t|^{d\nu} f_\pm ( g |t|^{-\phi}), 
\label{freeen3}
\end{equation} 
where $t\equiv (T-T^*)/T^*$, $g\equiv p-p^*$, and $\phi$ is the crossover exponent.

These results give rise to the following predictions for the
FSS behavior around $T^*$, $p^*$. Let us consider a
RG invariant quantity $R$, such as $R_\xi\equiv \xi/L$, $U_4$,
$U_{22}$, which are defined in the Appendix, and its derivative $R'$
with respect to $\beta\equiv 1/T$.  In general, in the FSS limit
$R$ obeys the scaling law
\begin{equation}
R = {\cal R}(u_1 L^{y_1}, u_2 L^{y_2}, \{u_i L^{y_i}\}),\quad i\ge 3.
\label{scalR}
\end{equation}
Neglecting the scaling corrections, that is terms vanishing
in the limit $L\to \infty$, close to the MGP we expect
\begin{equation}
R = R^* + b_{11} u_1 L^{y_1} + b_{21} u_2 L^{y_2} + \ldots.
\label{scalR1}
\end{equation}
which is valid as long as $u_1 L^{y_1}$ is small.
Along the N-line, the scaling field $u_2$ vanishes, so that 
we can write 
\begin{equation}
R_N = R^* + b_{11} u_1 L^{y_1} + \ldots,
\label{RGinvsca}
\end{equation}
where the subscript $N$ indicates that $R$ is restricted to the 
N-line. Let us now consider the derivative of $R$ with respect to $\beta$. 
Differentiating Eq.~(\ref{scalR1}), we obtain 
\begin{equation}
R' = b_{11} u'_1 L^{y_1} + b_{21} u'_2 L^{y_2} + \cdots 
\label{RGinvdsca1}
\end{equation}
If Eq.~(\ref{u1sf}) holds, then $u'_1=0$, so that
\begin{equation}
R' = b_{21} u'_2 L^{y_2} + \cdots 
\label{RGinvdsca2}
\end{equation}
This result gives us a method to verify the conjecture 
of Ref.~\onlinecite{LH-89}: once $y_1$ has been determined from the scaling 
behavior of a RG invariant ratio close to the MGP, it is enough to check
the scaling behavior of $R'$. If $R'$ scales as $L^x$ with $x<y_1$, 
the conjecture is confirmed and $x$ provides 
an estimate of $y_2$. 

Finally, we consider the magnetic susceptibility. Along the N-line it
behaves as 
\begin{equation}
\chi_N = e L^{2-\eta}\left( 1 + e_1 u_1 L^{y_1} + \cdots\right).
\label{RGinvchi}
\end{equation}
Note that there is only one $\eta$ exponent which characterizes the
critical behavior of both the magnetic and overlap correlation
functions,\cite{LH-88} since they are equal along the N-line,
see Eq.~(\ref{Grel}).

\section{Results} \label{sec3}

In the following we present a FSS analysis of high-statistics
MC data along the N-line close to the MGP.  We
performed MC simulations for lattice sizes 
$L=8,12,16,24,32$, taking periodic boundary conditions. 
We used a standard
Metropolis algorithm and multispin coding (details can be
found in Ref.~\onlinecite{HPPV-07-pmj}). 
Most of the simulations correspond to values of $p$ in the range 
$0.7680\le p \le 0.7685$, i.e. very close to the MGP, which, as we show below,
 is located at $p^* = 0.76820(4)$: typically, we considered 6 values of $p$
in this range for each value of $L$. To obtain small statistical errors,
we generated a large number of samples: $2\times 10^5$ for $L\le 16$,
$10^5$ for $L=24$, and $4\times 10^4$ for $L=32$. Because of the 
long equilibration times, for each sample we performed a large number 
of Metropolis sweeps; for $L=16$, 24, 32, the number of sweeps is 
$10^6$, $8\times 10^6$, and $5\times 10^7$, respectively. To guarantee
equilibration, typically 30\% of the data were discarded (but, for $L=32$, 
we discarded 50\% of the data). All MC data are available 
on request. Below we report the 
results of the analyses: in Sec.~\ref{sec3.1} we consider the static 
exponents, while in Sec.~\ref{sec3.2} we focus on the dynamics.

\subsection{Static exponents} \label{sec3.1}

\begin{figure*}[tb]
\centerline{\psfig{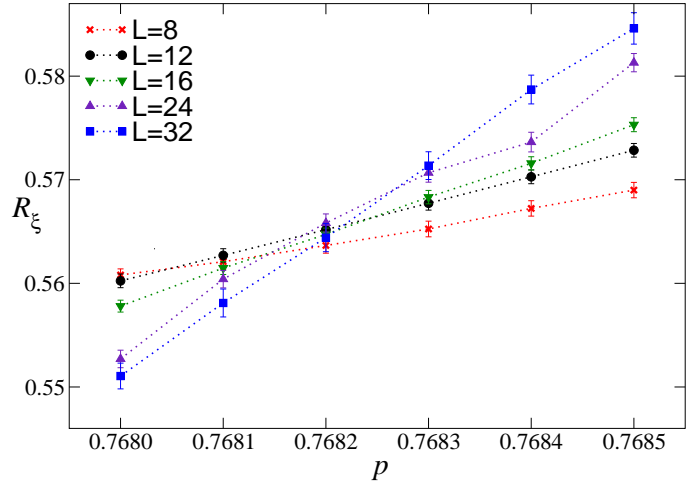}}
\vspace{4mm}
\centerline{\psfig{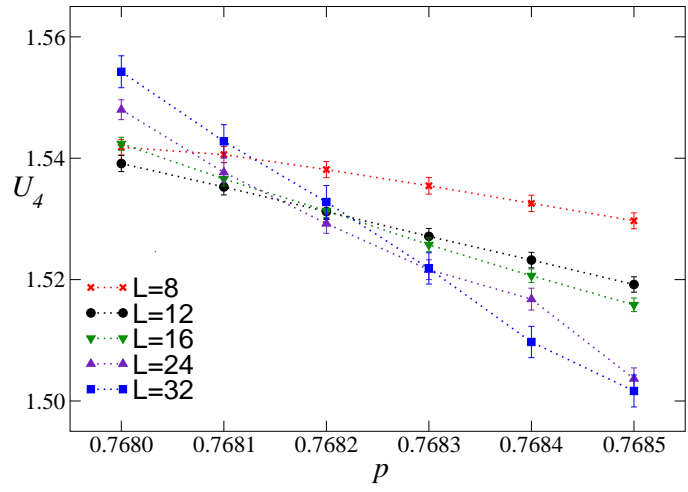}}
\vspace{4mm}
\centerline{\psfig{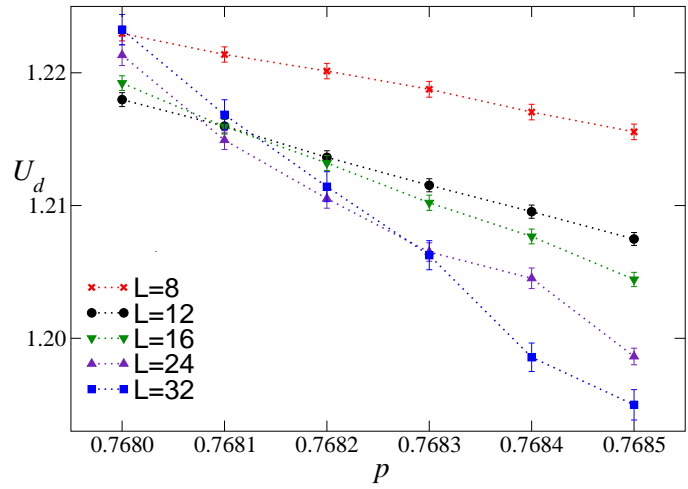}}
\vspace{2mm}
\caption{
MC data of $R_\xi\equiv \xi/L$, $U_4$, and $U_d$ vs $p$.
}
\label{rgifig}
\end{figure*}

MC estimates of the RG invariant quantities $R_\xi$, $U_4$, and $U_d$ 
along the N-line are shown in Fig.~\ref{rgifig}.
There is clearly a crossing point at $p \approx 0.7682$, 
which provides a first rough estimate of the location of the MGP point.
In order to estimate precisely $p^*$, $T^*$, and $y_1$ we fit the renormalized 
couplings $R$ close to the MGP to
\begin{equation}
R = R^* + a (\beta - \beta^*) L^{y_1},
\end{equation}
keeping $R^*$, $\beta^*$, and $y_1$ as free parameters. Note that this functional
form relies on the property that $u_2 = 0$ along the N-line. 
Otherwise, an additional term of the form $(\beta - \beta^*) L^{y_2}$ 
should be added. We also neglect scaling corrections that behave
as $c L^{y_3}$ with $y_3 < 0$. Indeed, since we only have data in a limited range 
of values of $L$, we are not able to include reliably a correction of this type.

\begin{table}
\begin{tabular}{lcccccc}
\hline\hline
 & $\chi^2$/DOF & $\beta^*$ & $y_1$ & $U_{22}^*$ & $R_\xi^*$ \\
\hline
$U_{22},R_\xi$ & 0.88 & 0.59910(2)&
     1.02(5) & $0.3180(3)$ &  0.5648(3) & \\
$U_{22},R_\xi,U_d$ & 1.43 & 0.59902(2)&
     1.02(4) & $0.3189(3)$ &  0.5640(4) & $U_{d}^*= 1.2137(3)$ \\
$U_{22},R_\xi,U_4$ & 0.62 & 0.59914(2)&
     1.01(5) & $0.3178(3)$ &  0.5656(4) & $U_{4}^*= 1.5302(6)$  \\
\hline\hline
\end{tabular}
\caption{Results of combined fits. The first fit uses all data with $L\ge 8$,
the last two fits only those  with $L\ge 12$. DOF is the number of degrees
of freedom of the fit.}
\label{Table1}
\end{table}

\begin{table}
\begin{tabular}{lrccc}
\hline\hline
 & $L_{\rm min}$ & $\chi^2$/DOF & $\beta^*$ & $x$  \\
\hline
$R_\xi,R_\xi'$ & 8 & 0.92 & 0.59905(3)& 0.600(2) \\
               &12 & 0.70 & 0.59912(3)& 0.609(4) \\
               &16 & 0.64 & 0.59910(4)& 0.604(7) \\
$U_{22},R_\xi'$& 8 & 0.69 & 0.59929(6)& 0.602(2) \\
               &12 & 0.55 & 0.59936(7)& 0.611(4) \\
               &16 & 0.46 & 0.59934(10)&0.607(7) \\
$R_\xi,U_4'$   & 8 & 2.06 & 0.59907(3)& 0.579(3) \\
               &12 & 0.71 & 0.59912(3)& 0.611(5) \\
               &16 & 0.60 & 0.59910(4)& 0.619(9) \\
$U_{22},U_4'$  & 8 & 1.60 & 0.59937(6)& 0.569(3) \\
               &12 & 0.55 & 0.59936(6)& 0.601(6) \\
               &16 & 0.43 & 0.59934(10)&0.607(10) \\
\hline\hline
\end{tabular}
\caption{Estimates of $x$. We report results obtained by analyzing 
simultaneously two different quantities and including only
data satisfying $L\ge L_{\rm min}$. DOF is the number of degrees of 
freedom of the fit.}
\label{Table2}
\end{table}

Fits that involve $R_\xi$ and $U_{22}$ have an acceptable $\chi^2$ 
even if we include all
data with $L\ge 8$: there is no evidence of scaling corrections.
On the other hand, in fits of $U_4$ or $U_d$ the data with $L=8$ must be discarded
to obtain a good $\chi^2$. To obtain more accurate estimates, we have performed combined fits
in which several RG invariant quantities are fitted together. The results are
reported in Table~\ref{Table1}. The dependence on the observables used in the fit
is reasonably small and allows us to estimate
\begin{eqnarray}
&&\beta^* = 0.5991(1), \label{estbetaN} \\
&&y_1 = 1.02(5). \label{esty1}
\end{eqnarray}
The errors take into account the variation of the estimates with the different
observables used in the fits (note that statistical errors are much smaller). 
Since scaling corrections are expected to differ in the different observables, 
this should allow us to take indirectly into account the scaling corrections.
We have then $T^* = 1/\beta^* = 1.6692(3)$, and, by using Eq.~(\ref{nishline}),
\begin{equation}
   p^* = 0.76820(4).
\label{pstar}
\end{equation}
In Table~\ref{Table1} we also report estimates 
of the critical value of the RG renormalized couplings. 
Note that $U_{22}^* \approx 0.318$, which is significantly higher than
the corresponding result for the RDIs universality class,
$U_{22}^* = 0.1479(6)$.\cite{HPPV-07} 
This indicates\cite{AH-96} that the violations of self-averaging
are much stronger at the MGP than along the PF transition line,
as of course should be expected. 

We consider now the derivative $R'$ of the RG invariant quantities with
respect to $\beta$. They have been determined by considering the 
connected correlations of $R$ and of the Hamiltonian. 
At the critical point, $R'$ is expected 
to behave as $L^x$ for large $L$, where $x = y_2$, if the argument of 
Ref.~\onlinecite{LH-89} holds; otherwise, one should have $x = y_1$. 
In order to determine $x$, we fit $\ln R'$ to 
\begin{equation}
\ln R' = a + x \ln L + b (\beta - \beta^*) L^{y_1},
\end{equation}
keeping $y_1$ fixed to $y_1 = 1.02(5)$. To avoid fixing $\beta_c$ 
we perform combined fits in which one derivative $R'_1$ and 
one RG coupling $R_2$ are fitted together. The results are reported in 
Table~\ref{Table2}. The $\chi^2$ of the fit is always good except
when we use $L_{\rm min} = 8 $ and $U_4'$. If we do not consider 
the corresponding results, all estimates of $x$ are close to $0.61$. 
Analyses of $R'_\xi$ are apparently stable with $L_{\rm min}$,
while those of $U'_4$ show a slight upward trend.
A reasonable final estimate is $x = 0.61(2)$, which takes into account all
results with their error bars.
This result is significantly different from $y_1$ and thus 
confirms the argument of Ref.~\onlinecite{LH-89}. Since $x < y_1$,
$x$ should be identified with $y_2$. Therefore, we obtain the estimates
\begin{equation}
y_2 = 0.61(2), \qquad\qquad \nu = {1\over y_2} = 1.64(5).
\end{equation}
The crossover exponent is therefore 
\begin{equation}
\phi = {y_1\over y_2} = 1.67(10).
\end{equation}

The same analysis used to estimate $y_2$ can be employed to determine $\eta$.
Instead of $\chi$, we consider the ratio $Z \equiv  \chi/\xi^2$, which has smaller statistical
errors. Since $Z\sim L^{-\eta}$ for $L\to\infty$ at the critical point, we fit
the MC data to 
\[
\ln Z = a - \eta \ln L + b (\beta - \beta^*) L^{y_1}.
\]
As before, we fix $y_1$ and perform combined fits of $\ln Z$ with 
a RG invariant coupling, considering only
data satisfying $L\ge L_{\rm min}$. Fits of $Z$ and $R_\xi$ give 
$\eta = -0.1155(6)$ and $-0.1154(9)$ for $L_{\rm min} = 8,12$; 
if we use $U_{22}$ instead of $R_\xi$, we obtain 
$\eta = -0.1134(7)$ and $-0.1131(9)$ for $L_{\rm min} = 8,12$.
The $L_{\rm min}$ dependence is small and results change only slightly
with the observable. We take as our final estimate
\begin{equation}
\eta = -0.114(3).
\end{equation}
Our FSS results significantly improve earlier results.
Ref.~\onlinecite{Singh-91} reports the computation and analysis of the
34th-order high-temperature (HT) series of some susceptibilities
\begin{equation}
\chi_{m,n} = {1\over V} \sum_{ij} [ \langle s_i s_j \rangle^m ]^n
\label{chimn}
\end{equation}
along the N-line, obtaining $p^*= 0.7656(20)$, $y_1=1.18(11)$,
$\phi\equiv y_1/y_2=1.85(14)$, $\eta=-0.10(2)$. These estimates are 
substantially consistent with ours. As a further check,
we reanalize the 34th-order
HT series reported in Ref.~\onlinecite{Singh-91}, by biasing the value
of the critical point with the MC estimate (\ref{estbetaN}).  Using biased
first-order integral approximants, see, e.g., Ref.~\onlinecite{CPRV-02} for details,
we obtain $(2-\eta)/y_1 = 2.08(7)$ from the series of $\chi_{11}$,
$(1-2\eta)/y_1 = 1.25(17)$ from the series of $\chi_{22}$, $3/y_1 = 3.03(14)$
from the series of the ratio $\chi_{11}^2/\chi_{22}$, and $(2-\eta-y_2)/y_1 =
2.70(9)$ from $v\partial \chi_{21}/\partial v$, from which we can derive the
estimates $y_1=0.99(5)$, $\phi\equiv y_1/y_2=1.6(3)$, and $\eta=-0.1(1)$,
which are in good agreement with our FSS results.  

Other results can be found in Refs.~\onlinecite{ON-87,Fisch-91,MB-98}; 
they are apparently less precise and not consistent with ours within the reported
errors. For example, we mention the recent estimates $p^*=0.7673(3)$
obtained by off-equilibrium MC simulations\cite{OI-98} and $p^*\approx 0.622$
obtained by a RG study.\cite{MB-98} Note that estimate ({\ref{pstar}) and the 
conjecture\cite{footnote} of Refs.~\onlinecite{TSN-05,Nishimori-07} 
allow us to find the location
of the multicritical point that occurs in the three-dimensional 
random-plaquette gauge model. We obtain $p^*_{\rm gauge} = 0.9650(1)$,
which is in agreement with, though much more precise than, the result
of Ref.~\onlinecite{OAIM-04}, $p^*_{\rm gauge} = 0.967(4)$.

\subsection{Model-A dynamic exponent $z$} \label{sec3.2}

\begin{figure*}[tb]
\centerline{\psfig{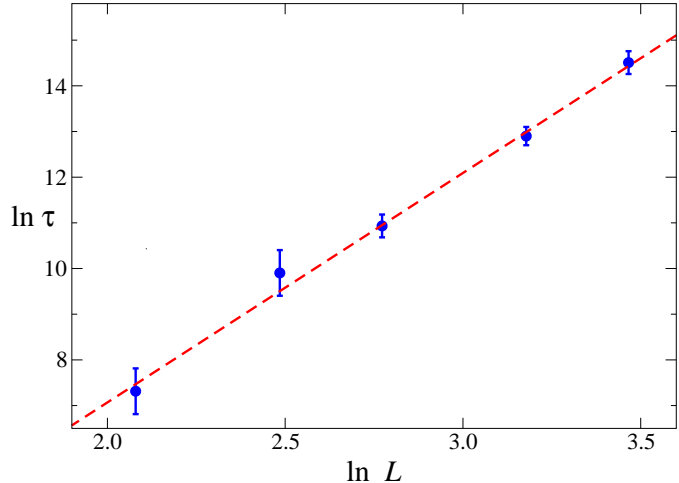}}
\caption{
Estimates of the exponential
autocorrelation time $\tau$ at the MGP vs $L$.
}
\label{taufig}
\end{figure*}

Finally, we present some results on the dynamic behavior of the Metropolis
algorithm, which represents a particular implementation of a relaxational
dynamics without conserved order parameters (model-A dynamics).\cite{HH-77} 
Note that at the MGP there is only one
dynamic exponent $z$ characterizing the relaxation of 
both the magnetic and the glassy critical modes,
since their autocorrelation functions
are strictly equal along the N-line.\cite{OI-98}
In Fig.~\ref{taufig} we show estimates of the exponential
autocorrelation time $\tau$ at the MGP  as extracted from the connected 
autocorrelation function of the magnetic susceptibility
\begin{equation}
G_\chi (t_1-t_2) \equiv \left[{ \langle \chi(t_1)\chi (t_2) \rangle_c }\right ].
\label{chitcorr}
\end{equation}
For large $L$ and $T=T^*$, $\tau$ is expected to scale as $L^z$, where $z$ is 
the dynamic critical exponent.
A linear fit of the MC results to $\ln \tau=a + b\ln L$ gives the estimate
$z=5.0(5)$, which is significantly larger than the value at the PF
transition line $z=2.35(2)$.\cite{HPV-07} Instead this estimate is close
to the value of $z$ obtained for the bimodal Ising spin-glass model,
$z = 5.7(2)$.\cite{PC-05}

We also determine the exponent $\lambda$ which describes the nonequilibrium
relaxation of the magnetization at $T_c$ from a starting 
configuration in which all spins are parallel.\cite{OI-98} 
Asymptotically, for $t\rightarrow \infty$, one
expects
\begin{equation}
M(t)\sim t^{-\lambda},\qquad \lambda={1+\eta\over 2z}, 
\label{mt}
\end{equation}
see, e.g., Ref.~\onlinecite{OI-98} and references therein.  
Our results lead to the estimate $\lambda=0.09(1)$,
which is perfectly consistent with the estimate~\cite{OI-98}
$\lambda=0.090(3)$ obtained in off-equilibrium MC simulations.

\section{Conclusions} \label{sec4}

In this paper we have considered the critical behavior close to the 
MGP which  is present in the phase diagram of the $\pm J$ model.
Our main results are the following:
\begin{itemize}
\item[(i)] We have obtained an accurate estimate of the location of the MGP:
$p^* = 0.76820(4)$, $\beta^* = 0.5991(1)$.  It is worth observing that
our estimate of $p^*$ is very close to  
the result~\cite{Hartmann-99} $p_c=0.778(5)$
for the location of the FG transition at $T=0$ and 
satisfies the rigorous inequality
$p_c \ge  p^*$ which follows from Eq.~(\ref{ineq}).
Our results show therefore that, even if 
the conjecture\cite{Nishimori-86,Kitatani-92} that the
FG transition line does not depend on the
temperature is not true, deviations are quite small.
\item[(ii)] We have verified the conjecture of Ref.~\onlinecite{LH-89}: 
the scaling 
field $u_1$ associated with the RG operator with the largest RG dimension
does not depend on the temperature.
\item[(iii)] We have determined the critical exponents 
$y_1 = 1.02(5)$, $y_2 = 0.61(2)$, $\nu\equiv 1/y_2=1.64(5)$, $\phi\equiv y_1/y_2 
= 1.67(10)$, and 
$\eta = -0.114(3)$.
\item[(iv)] We have determined the dynamic critical exponent $z$ associated
with the model-A dynamics, obtaining $z = 5.0(5)$. 
\end{itemize}
Our results are significantly more precise than those obtained in 
previous works.\cite{ON-87,Fisch-91,Singh-91,SA-96,MB-98,OI-98} They can be 
used to explain experiments on materials containing
both ferromagnetic and antiferromagnetic ions. An example is 
Fe$_x$Mn$_{1-x}$TiO$_3$, which shows Ising behavior for $x=1$ and $x=0$, a 
PG transition for $0.38\lesssim x \lesssim 0.58$ 
and a PF transition for $0 < x \lesssim 0.38$ and 
$0.58\lesssim x < 1$.\cite{YMAI-89,AKI-93} The MGP should be 
located at $x\approx 0.38$ and at $x\approx 0.58$. Close to these values our results
apply.

\section*{Acknowledgments}

The MC simulations
have been done at the Computer Laboratory of the Physics Department of Pisa.

\appendix

\section{Notations}

Setting
\begin{equation}
G_k(x) \equiv [ \langle \sigma_0 \,\sigma_x \rangle^k ],
\label{twopk}
\end{equation}
where the angular parentheses and the brackets indicate respectively
the thermal average and the quenched average over $J_{xy}$,
the magnetic and overlap correlation functions
are given respectively by $G_1(x)$ and $G_2(x)$.
Along the N-line, cf. Eq.~(\ref{nishline}), $G_1(x)=G_2(x)$.

We define the magnetic susceptibility $\chi\equiv \sum_x G_1(x)$ and the
correlation length $\xi$
\begin{equation}
\xi^2 \equiv {\widetilde{G}_1(0) - \widetilde{G}_1(q_{\rm min}) \over 
          \hat{q}_{\rm min}^2 \widetilde{G}_1(q_{\rm min}) },
\end{equation}
where $q_{\rm min} \equiv (2\pi/L,0,0)$, $\hat{q} \equiv 2 \sin q/2$, and
$\widetilde{G}_1(q)$ is the Fourier transform of $G_1(x)$.  We also consider
quantities that are invariant under RG transformations in the critical limit.
Beside the ratio
\begin{equation}
R_\xi \equiv \xi/L,
\label{rxi}
\end{equation}
we consider the quartic cumulants $U_4$, $U_{22}$, and $U_d$ defined by
\begin{eqnarray}
&& U_{4}  \equiv { [ \mu_4 ]\over [\mu_2]^{2}}, 
\label{cumulants}\\
&&U_{22} \equiv  {[ \mu_2^2 ]-[\mu_2]^2 \over [\mu_2]^2},
\nonumber \\
&&U_d \equiv U_4 - U_{22},
\nonumber
\end{eqnarray}
where
\begin{eqnarray}
\mu_{k} \equiv \langle \; ( \sum_x \sigma_x\; )^k \rangle \; .
\end{eqnarray}
Analogous quantities $R^o_\xi$, $U^o_4$, $U^o_{22}$, and $U^o_d$ 
can be defined by using the overlap variable
$q_x\equiv \sigma_x^{(1)}\sigma_x^{(2)}$, where the superscripts indicate two independent
configurations for given disorder. 
Using Eq.~(\ref{nishrel}), one can easily check that along the N-line 
$R_\xi=R^o_\xi$ and $U_4=U^o_4$. This implies that also their fixed-point
values are the same at the MGP.

Finally, we consider the derivatives
\begin{equation}
R'_\xi\equiv {d R_\xi\over d\beta},\qquad
U'_4\equiv {d U_4\over d\beta},
\label{derivatives}
\end{equation}
which can be computed by measuring appropriate expectation values
at fixed $\beta$ and $p$.


\begin{thebibliography}{99}

\bibitem{EA-75}
S. F. Edwards and P. W. Anderson,
J. Phys. F {\bf 5}, 965 (1975).

\bibitem{IATKST-86}
A. Ito, H. Aruga, E. Torikai, M. Kikuki, Y. Syono, and
H. Takei, Phys. Rev. Lett. {\bf 57}, 483 (1986).

\bibitem{GSNLAI-91}
K. Gunnarsson, P. Svedlindh, P. Nordblad, L. Lundgren,
H. Aruga, and A. Ito,
Phys. Rev. B {\bf 43}, 8199 (1991).

\bibitem{NN-07}
S. Nair and A. K. Nigam,
Phys. Rev. B {\bf 75}, 214415 (2007).

\bibitem{Hukushima-00}
K. Hukushima, J. Phys. Soc. Japan {\bf 69}, 631 (2000).

\bibitem{HPPV-07-pmj}
M. Hasenbusch, F. Parisen Toldin, A. Pelissetto, and E. Vicari,
Phys. Rev. B {\bf 76}, (2007) 
[arXiv:cond-mat/0704.0427].

\bibitem{HPPV-07}
M. Hasenbusch, F. Parisen Toldin, A. Pelissetto, and E. Vicari,
J. Stat. Mech.: Theory Expt. P02016 (2007). 

\bibitem{PV-02}
A. Pelissetto and E. Vicari,
Phys. Rept. {\bf 368}, 549 (2002).

\bibitem{CPPV-04}
P. Calabrese, P. Parruccini, A. Pelissetto, and E. Vicari,
Phys. Rev. E {\bf 69}, 036120 (2004).

\bibitem{Aharony-76} 
A.~Aharony, in
{\em Phase Transitions and Critical Phenomena},
Vol.\ 6, edited by C.~Domb and M.S.~Green
(Academic Press, New York, 1976), p. 357.

\bibitem{CPRV-02}
M. Campostrini, A. Pelissetto, P. Rossi, and E. Vicari,
Phys. Rev. E {\bf 65}, 066127 (2002).

\bibitem{meanfield}
G. Toulouse, J. Physique Lettres {\bf 41}, 447 (1980).

\bibitem{KKY-06}
H. Katzgraber, M. K\"orner, and A. P. Young,
Phys. Rev. B {\bf 73}, 224432 (2006).

\bibitem{KR-03}
N. Kawashima and H. Rieger,  in 
{\em Frustrated Spin Systems}, edited by H.T. Diep
(World Scientific, Singapore, 2004); cond-mat/0312432.

\bibitem{GHDB-85}
A. Georges, D. Hansel, P. Le Doussal, and 
J. Bouchaud, J. Phys. (Paris) {\bf 46}, 1827 (1985).

\bibitem{LH-88}
P. Le Doussal and A. B. Harris,
Phys. Rev. Lett. {\bf 61}, 625 (1988).

\bibitem{LH-89}
P. Le Doussal and A. B. Harris,
Phys. Rev. B {\bf 40}, 9249 (1989).

\bibitem{Nishimori-81}
H. Nishimori, Prog. Theor. Phys. {\bf 66}, 1169 (1981).

\bibitem{Nishimori-book}
H. Nishimori, {\em Statistical Physics of Spin Glasses and 
Information Processing: An Introduction}\/
(Oxford University Press, Oxford, 2001).

\bibitem{Nishimori-86}
H. Nishimori, J. Phys. Soc. Japan {\bf 55}, 3305 (1986).

\bibitem{Kitatani-92}
H. Kitatani, J. Phys. Soc. Japan {\bf 61}, 4049 (1992).

\bibitem{WHP-03}
C. Wang, J. Harrington, and J. Preskill,
Ann. Phys. {\bf 303}, 31 (2003).

\bibitem{AH-04}
C. Amoruso and A. K. Hartmann,
Phys. Rev. B {\bf 70}, 134425 (2004).

\bibitem{PHP-06}
M. Picco, A. Honecker, and P. Pujol,
J. Stat. Mech.: Theory Expt. P09006 (2006).

\bibitem{Kitatani-94}
It can be shown rigorously that the N-line never intersects the 
spin-glass phase, H. Kitatani, J. Phys. Soc. Japan 
{\bf 63}, 2070 (1994). Since we must also have $p_{FG}\le p^*$,
the mixed phase, if it exists, should be 
confined to the region below the N-line and on the left of 
the line $p = p^*$ (see Fig.~\ref{phdia}). 

\bibitem{SK-75}
D. Sherrington and S. Kirkpatrick,
Phys. Rev. Lett. {\bf 35}, 1792 (1975).

\bibitem{CKR-05}
T. Castellani, F. Krzakala, and F. Ricci Tersenghi,
Eur. Phys. J. B {\bf 47}, 99 (2005).

\bibitem{Hartmann-99}
A. K. Hartmann, Phys. Rev. B {\bf 59}, 3617 (1999).

\bibitem{KM-02}
F. Krzakala and O.C. Martin, 
Phys. Rev. Lett. {\bf 89}, 267202 (2002).

\bibitem{HH-77}
P. C. Hohenberg and B. I. Halperin,
Rev. Mod. Phys. {\bf 49}, 435 (1977).

\bibitem{ON-87}
Y. Ozeki and H. Nishimori, 
J. Phys. Soc. Japan {\bf 56}, 1568 (1987);
J. Phys. Soc. Japan {\bf 56}, 3265 (1987).

\bibitem{Fisch-91}
R. Fisch, Phys. Rev. B {\bf 44}, 652 (1991).

\bibitem{Singh-91}
R. R. P. Singh, 
Phys. Rev. Lett. {\bf 67}, 899 (1991).

\bibitem{SA-96}
R. R. P. Singh and J. Adler,  Phys. Rev. B {\bf 54}, 364 (1996).

\bibitem{MB-98}
G. Migliorini and A. N. Berker,
Phys. Rev. B {\bf 57}, 426 (1998).

\bibitem{OI-98}
Y. Ozeki and N. Ito, J. Phys. A {\bf 31}, 5451 (1998).

\bibitem{AH-96}
A. Aharony and A. B. Harris, 
Phys. Rev. Lett. {\bf 77}, 3700 (1996).

\bibitem{footnote}
Note that the duality relations reported in Ref.~\onlinecite{TSN-05}
are not rigorous. Numerical results are generically consistent
(see Table I in Ref.~\onlinecite{Nishimori-07}), even though 
tiny discrepancies have been observed in several cases.

\bibitem{TSN-05}
K. Takeda, T. Sasamoto, and H. Nishimori, 
J. Phys. A {\bf 38}, 3751 (2005).

\bibitem{Nishimori-07}
H. Nishimori, J. Stat. Phys. {\bf 126}, 977 (2007).

\bibitem{OAIM-04}
T. Ohno, G. Arakawa, I. Ichinose, and T. Matsui,
Nucl. Phys. B {\bf 697}, 462 (2004).

\bibitem{HPV-07}
M. Hasenbusch, A. Pelissetto, and E. Vicari, 
in preparation.

\bibitem{PC-05}
M. Pleimling and I. A. Campbell, Phys. Rev. B {\bf 72},
184429 (2005).

\bibitem{YMAI-89}
H. Yoshizawa, S. Mitsuda, H. Aruga, and A. Ito,
J. Phys. Soc. Jpn. {\bf 58}, 1416 (1989).

\bibitem{AKI-93}
H. Aruga Katori and A. Ito,
J. Phys. Soc. Jpn. {\bf 62}, 4488 (1993).




\end{thebibliography}
\end{document}